\documentclass[reprint,pra,superscriptaddress,twocolumn,showpacs]{revtex4-1}
\usepackage{graphicx}
\usepackage{amsmath}
\usepackage{amsfonts}
\usepackage{bm}
\usepackage{color}
\usepackage{cancel}

\usepackage[plainpages=false,pdfpagelabels,colorlinks=true,linkcolor=red,urlcolor=blue,citecolor=blue,pdftitle={Title},pdfauthor={},pdfdisplaydoctitle=true,pdfduplex=DuplexFlipLongEdge]{hyperref}

\definecolor{darkred}{rgb}{0.90,0.2,0.2}

\begin{document}
\title{
Nature of Bosonic Excitations revealed by  high--energy charge carriers }

\author{Jan \surname{Kogoj}}
\affiliation{J. Stefan Institute, 1000 Ljubljana, Slovenia}

\author{Marcin \surname{Mierzejewski}}
\affiliation{Institute of Physics, University of Silesia, 40-007 Katowice, Poland}

\author{Janez \surname{Bon\v{c}a}}
\affiliation{J. Stefan Institute, 1000 Ljubljana, Slovenia}
\affiliation{Faculty of Mathematics and Physics, University of Ljubljana, 1000
Ljubljana, Slovenia}

\begin{abstract}

We address a long standing problem concerning the origin  of  bosonic excitations that strongly interact with charge carriers. We show that the time-resolved pump--probe  experiments are capable  to distinguish  
between regular bosonic degrees of freedom, e.g. phonons, and the hard-core  bosons, e.g.,  magnons. The  ability of phonon degrees of freedom to absorb essentially unlimited amount of energy renders relaxation dynamics nearly independent  on the absorbed energy or the fluence.  In contrast, the hard core effects pose limits on the density of energy stored in the bosonic subsystems  resulting in a substantial dependence of the relaxation time on the fluence and/or excitation energy.  
Very similar effects can be observed also in a different setup when the system is driven by multiple pulses of equal energy.  
\end{abstract}
\maketitle

	Solids are complex objects with different degrees of freedom, hence the charge carriers are simultaneously coupled to various types of bosonic excitations like phonons, magnons, plasmons or others.   The longstanding problem in the studies on strongly correlated systems is to single out the strongest,  and  thus probably  the most relevant interaction.  
	However for many important materials, including the unconventional superconductors, we are still seeking the answer to a more modest question, whether the strongest coupling {of charge carriers} is to phonons or to some kind of magnetic excitations.  The essential qualitative difference between these excitations is that the latter ones are hard-core (HC) quasiparticles, i.e. their spatial density is limited typically by one boson per lattice site.  It is rather clear that  the HC effects become important first in the vicinity of this bound, i.e. when the energy density is of the order of the frequency of the bosonic excitations.  However, it might be impossible to heat up the entire system to such  high  energies since, e.g., the corresponding temperature may be too close to the melting point.  A very promising solution is to excite only targeted degrees of freedom.  Directly  after such excitation, the total system is far from  equilibrium since various degrees of freedom may have very different temperatures (energies to be more precise).   

	This solution is utilised  in the recent time resolved photoemission and optical spectroscopies, where one studies ultrafast relaxation of a few highly excited charge carriers \cite{okamoto2010,gadermaier10,cortes11,dalconte12,gadermaier14,novelli14,rameau14,rameau15,dalconte15}.   These carriers lower high initial kinetic energy (of the order of 1-2eV) in a narrow time-window, by emitting many of most strongly coupled bosons. In the vicinity of the excited charge carriers, the concentration of emitted bosons may be high enough so that the HC effects become visible.

	In the majority of the recent theoretical papers,  the ultrafast relaxation of highly excited carriers has been discussed separately for the couplings to phonons \cite{golez_prl2012,defilippis12,matsueda12,kennes10,kemper13,sentef13,werner2013,baranov14,kemper14,werner15,aoki2015,lev2015,sayyad15,mishchenko15,rizzi16}  
and to the magnetic excitations ~\cite{golez2014,iyoda14,eckstein14b,golez2016}.  On the one hand, the electron-phonon coupling is usually described within the Holstein model by two independent parameters: the dimensionless coupling strength, $\lambda$ and  the phonon frequency, $\omega$. The electron-phonon interaction  is efficient in any dimension, hence it has been mostly studied in the simplest case of the one-dimensional (1D) systems.  On the other hand, the coupling to the magnetic excitations is studied mainly within the Hubbard or the t-J models. Each of the latter models contains a single free parameter  ($U/t$ or $J/t$) which encodes the coupling strength as well as the frequency of the excitations.   Since the spin and the charge degrees of freedom are separated  in the 1D systems, relaxation due to magnetic excitations should be studied at least in two dimensions. Due to the essential differences between the studied models, it is difficult to specify the distinctive qualitative features of both  mechanisms. In the present work we fill this important gap. We compare the relaxation of charge carriers in two models, chosen such that all the emerging differences are solely due to  HC effects of the bosonic excitations.
{We show that if the relaxation is due to the coupling to HC  bosons, then the relaxation time  shows pronounced dependence on the excitation energy and/or 
the density of excited carriers. The opposite holds true for coupling  do standard bosonic degrees of freedom.} 

{\it Model}. %
We consider two models on a one-dimensional ring with $L$ sites, each containing a single electron. The first one is the  Holstein model (HM) while the second is a HC boson model (HCM) where phonon degrees of freedom are replaced by hard core bosons. Hamiltonians of both models have a very similar structure: 
\begin{equation} 
H_{\rm sys} = H_{\rm kin} + H_{\rm bos} + H_{\rm int},
\label{ham}
\end{equation}
where individual part of $H_{\rm sys}$ are as follows:
\begin{eqnarray}
H_{\rm kin} &=& -t_{0}\sum_{j} \left({e^{i \phi(t)}} c_{j}^{\dagger}  c_{j+1} + \mbox{h.c.} \right)\nonumber \\
H_{\rm bos} &=& \omega \sum_{j} b_{j}^\dagger  b_{j},\label{ham_parts}\\
H_{\rm int} &=& -g \sum_{j} c_{j}^{\dagger}c_{j} (b_{j}^\dagger + b_{j}),\nonumber \\
\end{eqnarray}
where $t_{0}$ is the  hopping amplitude, and  $c_{j}$ is a fermion annihilation operator on site $j$.
Operators $b_j$ represent either phonon or HC boson field annihilation operator. 
There is at most one HC boson per site, hence $b^{\dagger}_i b^{\dagger}_i =0$.  This restriction shows up  in specific commutation relations $[b_i,b^{\dagger}_j ]=\delta_{ij}(1-2 b^{\dagger}_i b_i)$ for the latter operators.
 For clarity of comparison we introduce identical dispersion-less frequency $\omega$  and coupling $g$ for phonons and HC bosons alike. {The $\phi(t)$ represents the phase gained by the electron as it hops between successive sites. It is used to pump energy into the system as described  in the last part of this work.} We measure time in units $\hbar/t_0$ and set $\hbar=1$. 

We solve both models using the  Lanczos-based diagonalization defined within  a limited functional space (LFS) to obtain the ground state as well as for the time evolution. The generation of the LFS efficiently selects states with different phonon configurations in the vicinity  of the electron thus enabling   numerically exact solution of the polaron problem \cite{bonca99,ku2002}, and it is well-suited  also  to describe polaron systems far from the equilibrium~\cite{lev2011_1,golez2012,golez_prl2012,lev2015}.  A detailed description of this numerical approach can be found in  Ref. \cite{lev2011_1}.


In the first part of this work we  start the time evolution from the free electron  wavefunction at a given wave number $k$, $\vert \psi_0(t=0)\rangle=c_k^\dagger\vert 0\rangle$ where $\vert 0\rangle$ represents vacuum state for electrons and bosons.   We choose the initial kinetic energy of the electron $ E_\mathrm{kin}(t=0)=-2t_0\cos(k)$ to be much larger than the ground-state   $E_\mathrm{kin}^\mathrm{GS}$  of the polaron. We then perform the  time evolution under  $H_\mathrm{sys}$, Eq.~\ref{ham}, to obtain the wavefunction $\vert \psi(t)\rangle$ of the system.   {This approach simulates   polaron formation starting from a free electron with variable (possibly very high) initial kinetic energy.} 

\begin{figure} 
\includegraphics[width=0.43\textwidth]{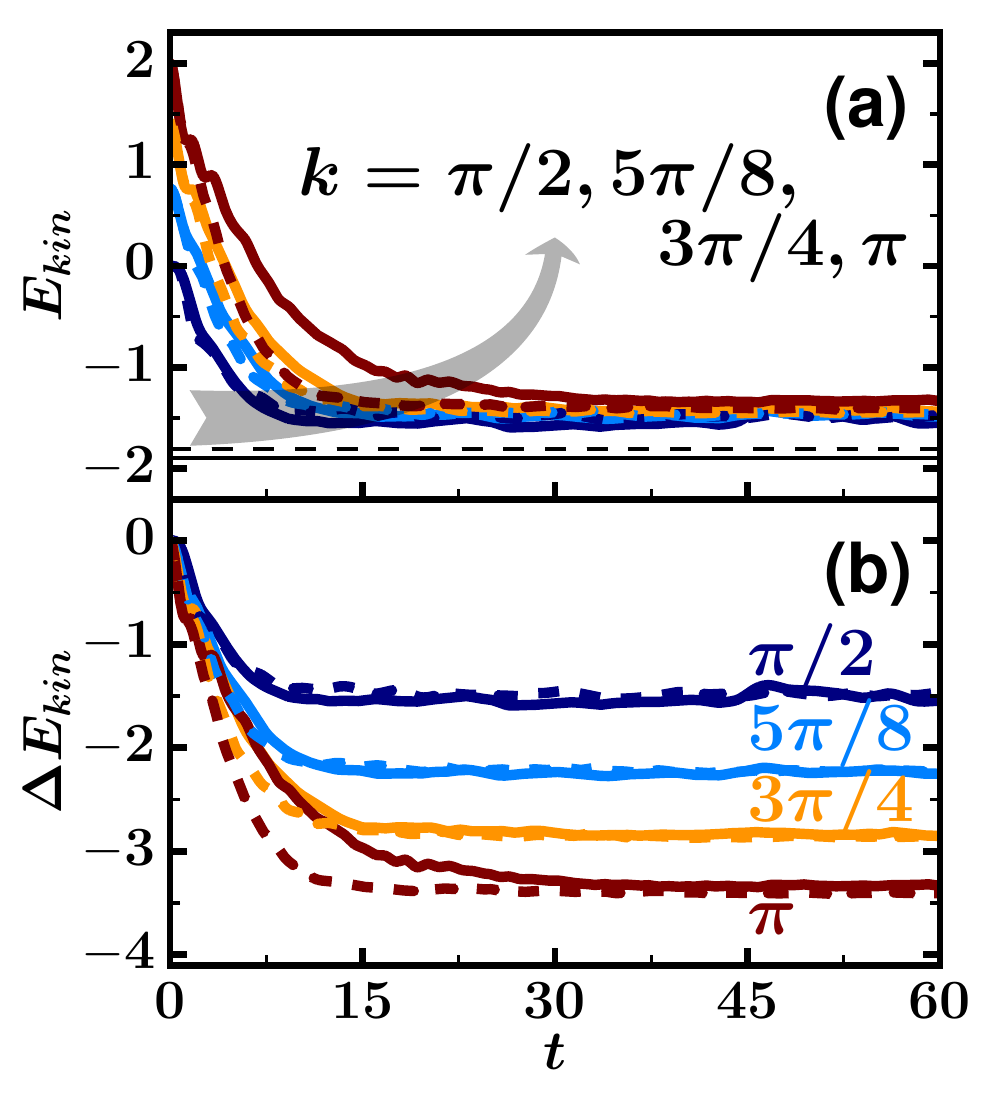}
\caption{$E_\mathrm{kin}(t)$ vs. time $t$ for different values of $k$ in (a) and $\Delta E_\mathrm{kin}(t)=E_\mathrm{kin}(t)-E_\mathrm{kin}(t=0)$ in (b). Dashed (full) lines are represent results for  HM (HCM). Other parameters of the Hamiltonian in Eq.~\ref{ham_parts} are $\omega=0.5$, $g=\sqrt{0.5}$ and the size of the system with periodic boundary conditions  is $L=16$. Unless  otherwise specified, identical  parameters were used in all subsequent figures.  { The phase is set to $\phi(t)=0$ in all figures but Fig~\ref{fig5}.} Thin horizontal lines present ground-state kinetic energies $E_\mathrm{kin}^\mathrm{GS}$.
}
\label{fig1}
\end{figure}
In Fig.~\ref{fig1}(a) we present comparison of $E_\mathrm{kin}(t)=\langle H_{\mathrm{kin}}(t)\rangle$ of HM and HCM for different values of initial $k$. In all cases  we observe a decrease of $E_\mathrm{kin}(t)$ towards a quasi steady state values $\bar E_\mathrm{kin}$ that remain consistently above their  respective ground-state values $E_\mathrm{kin}^\mathrm{GS}$.  
We should stress that the total energy of the system $ E_\mathrm{sys}$ remains constant during the time evolution and it equals the initial value of the kinetic energy, {\it i.e.} $ E_\mathrm{sys}=E_\mathrm{kin}(t=0)$.  The decrease of $E_\mathrm{kin}(t)$ is thus intimately connected to the increase of the phonon or HC boson energy. The main difference between the models under  consideration is that phonon degrees of freedom can absorb in principle an infinite amount of energy while HC bosons can absorb at most $\omega$ of energy per site. 

To facilitate further comparison of relaxation dynamics between two different models,  we have shifted all initial values of $E_\mathrm{kin}$ to $E_\mathrm{kin}(t=0)=0$, see Fig.~\ref{fig1}(b). For small $k$ , {\it i.e.} $k=\pi/2$ and $5\pi/8$,  relaxation dynamics of HM and HCM are nearly indistinguishable. For larger values  of $k\geq 3\pi/4$, however,  relaxation of the HM  seems to be substantially  faster than that of the HCM. {Nevertheless, the quasi-steady state values $\bar E_\mathrm{kin}$ are nearly indistinguishable between the two models. }
  Even though the relaxation process is not strictly exponential, we nevertheless found very reasonable exponential fits to the data, presented in Fig.~\ref{fig1}(b). 
\begin{figure} 
\includegraphics[width=0.43\textwidth]{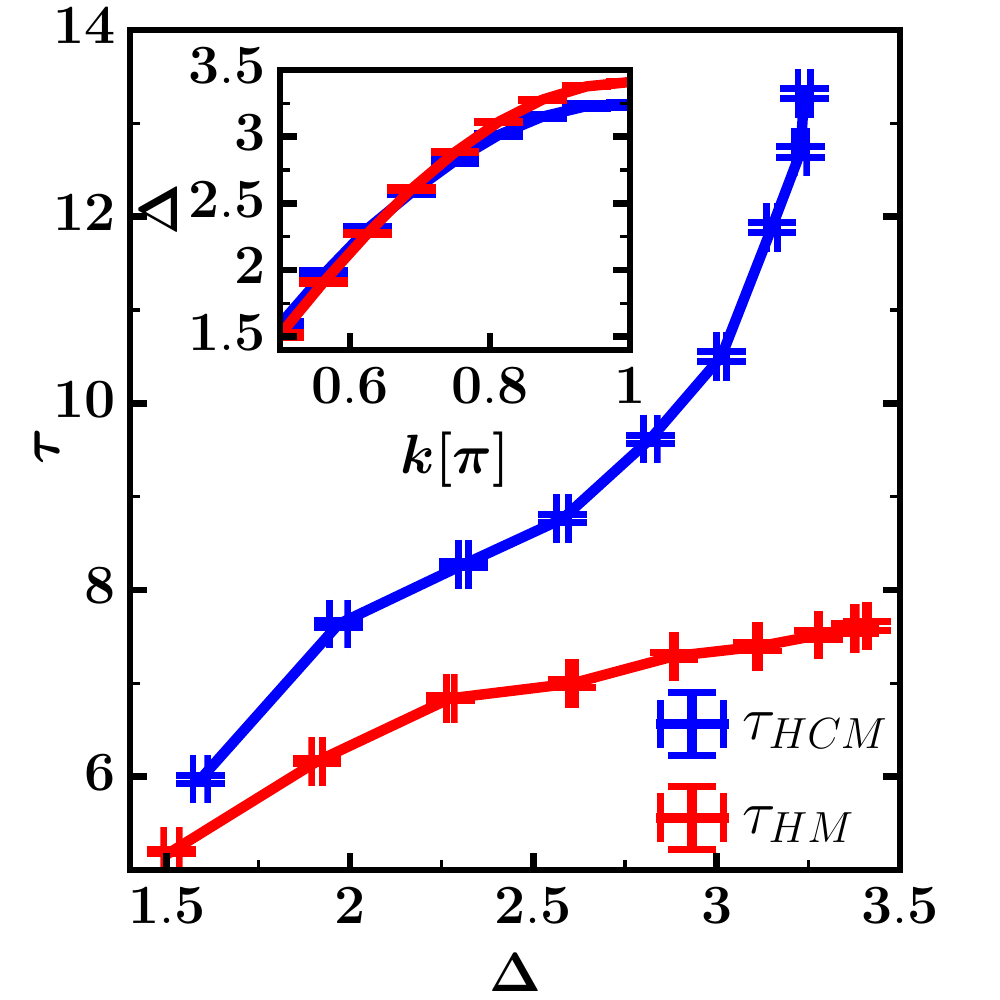}
\caption{
Relaxation time $\tau_\mathrm{HM}$ ($\tau_\mathrm{HCM}$) for HM (HCM) vs. quench energy 	$\Delta$. $\tau$ were  extracted from data presented in Fig.~\ref{fig1}(b) using the following analytical form $\Delta E_\mathrm{kin}(t) = A\exp{[-t/\tau]}+B$. 
In the inset  we show the dependence of  $\Delta$ on the wave number of the initial free electron wavefunction $k$.  }
\label{fig2}
\end{figure}

In Fig.~\ref{fig2} we present the central result of this work, that is, the comparison of relaxation times $\tau_\mathrm{HM}$ and $\tau_\mathrm{HCM}$ of the HM and HCM, respectively as a function of the quench energy that is defined as a difference between the initial kinetic energy of the electron and the average kinetic energy in the quasi steady state:  $\Delta =E_\mathrm{kin}(t=0)-\bar E_\mathrm{kin}$. 
{ We found a relatively  weak dependence of relaxation times $\tau_\mathrm{HM}$  on $\Delta $ in particular, there is no abrupt raise of  $\tau_\mathrm{HM}$ at large values of $\Delta$. In contrast,  much more pronounced  dependence on $\Delta $ is found in the case of $\tau_\mathrm{HCM}$, see Fig.~\ref{fig2}. We observe a sharp up-turn of $\tau_\mathrm{HCM}$ for larger values of $\Delta $ signalling a significant slowing down of the relaxation process in the HCM.  In contrast, at small $\Delta\sim 1.5$,   $\tau_\mathrm{HCM}$ approaches $\tau_\mathrm{HM}$. We should also stress, that relaxation times for $\Delta \lesssim 2$ become less reliable since relaxation processes evolve over a smaller number of eigenstates.  The dependence of $\Delta$ on the wave vector $k$ is presented in the insert of Fig.~\ref{fig2}.}

\begin{figure} 
\includegraphics[width=0.43\textwidth]{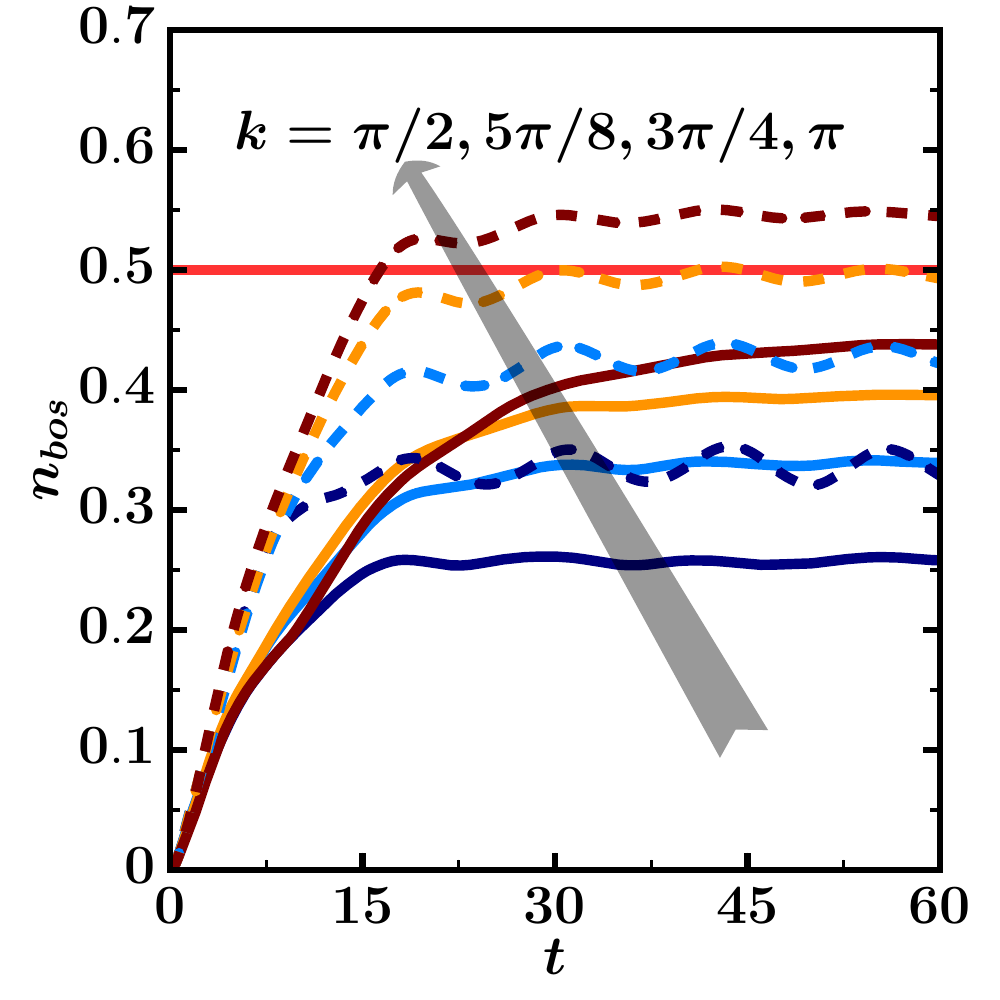}
\caption{ Average number of bosonic excitations per-site $n_\mathrm{bos}$ vs. $t$ for different values of $k$ corresponding to different $\Delta$ as depicted in the inset of Fig.~\ref{fig2}. Dashed (full) lines represent results for the HM (HCM). The horizontal line indicates the infinite--temperature value of  $n_\mathrm{bos}$ in the HCM. 
}
\label{fig3}
\end{figure}
To gain a deeper understanding of the different relaxation processes in models under the investigation,  we computed the average number of bosonic excitations per site, given by $n_\mathrm{bos}=\langle\psi(t)\vert \sum_i b^\dagger_i b_i\vert\psi(t)\rangle/L$, shown in Fig.~\ref{fig3}. We first note that   there is no upper bound on   $n_\mathrm{bos}$ in the HM case. In contrast,  there is a formal upper bound $n_\mathrm{bos}  \le 1$ in the HCM while  twice smaller concentration  $n_\mathrm{bos} = 0.5$ is  reached only  in the limit of infinite temperature. In both models we observe an increase of $ n_\mathrm{bos}$ at longer times as the $k$ and with it $\Delta $ increase. While  $ n_\mathrm{bos}$ in the HM keeps increasing with increasing $k$, we observe signs of saturation in the HCM case as  $ n_\mathrm{bos}$ moves closer to its infinite--temperature value $ n_\mathrm{bos}=0.5$ at maximal quench energy reached at $k=\pi$.  In the HM we observe pronounced oscillations in the long-time regime with the time-period $t_p\sim 2\pi/\omega$ that are not reflected in oscillations of $E_\mathrm{kin}(t)$ since they are compensated by oscillations in the {interaction  energy $E_\mathrm{int}=\langle H_\mathrm{int} \rangle $}, not shown. { We observe also    that $n_\mathrm{bos}$ in the HM systematically substantially exceeds $n_\mathrm{bos}$ in the HCM. Taking also into account that total energies of both models together with nearly equal  $\bar E_\mathrm{kin}$, see Fig.~\ref{fig1}, a large difference in $n_\mathrm{bos}$ represents a seeming  contradiction. Nevertheless, the excess in bosonic energy is compensated by a lower interaction energy $E_\mathrm{int}$ (not shown) in the HM.}
\begin{figure} 
\includegraphics[width=0.42\textwidth]{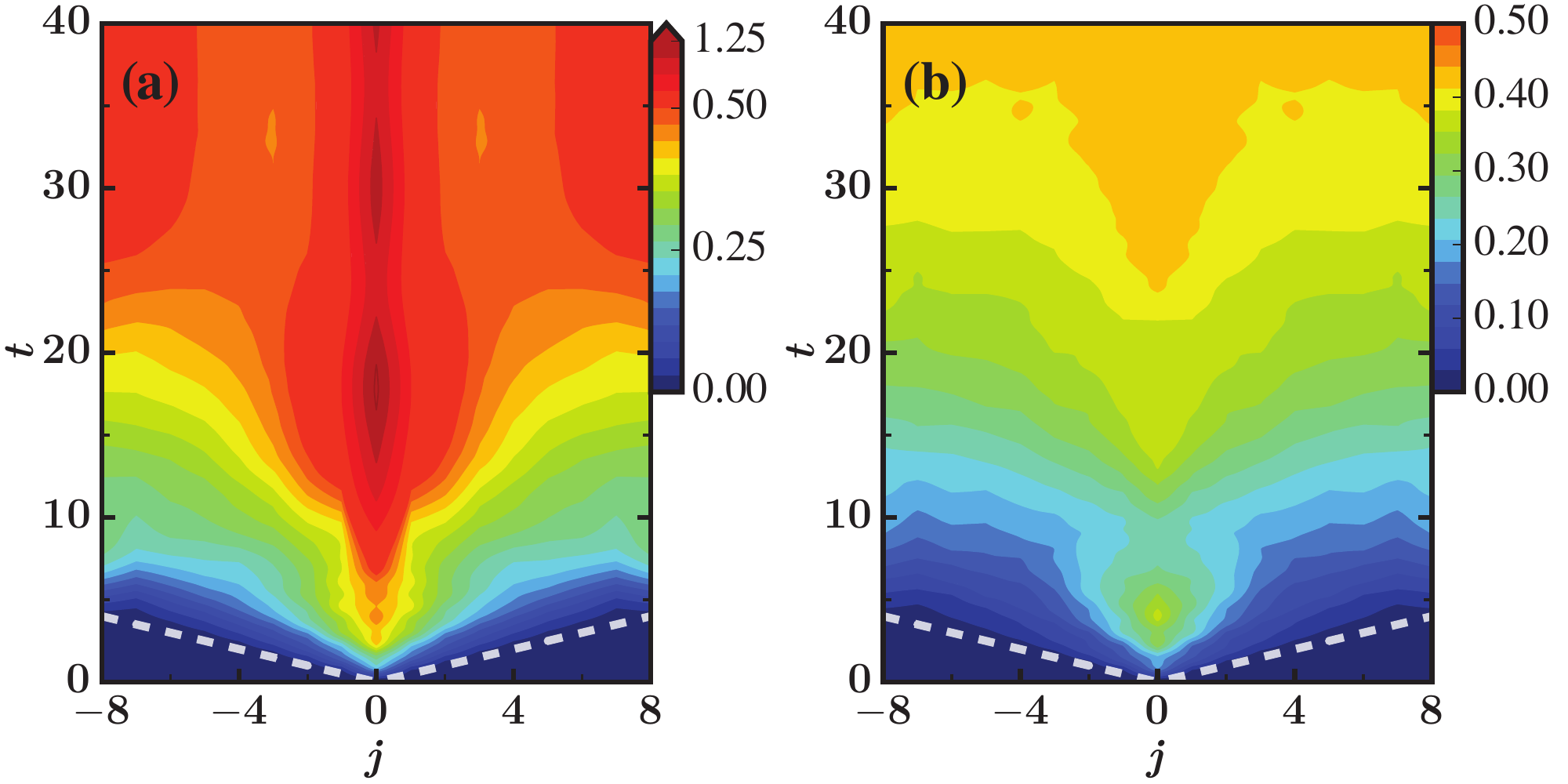}
\caption{ Density plot of the spread of bosonic excitations $\gamma(j,t)$ for the HM in a) and HCM in b) for $k=\pi$.  Dashed white lines represent the Lieb Robinson bound as described in the text. To facilitate a straightforward comparison  between models, identical colour coding  was used in both figures. 
}
\label{fig4}
\end{figure}

We next present in Fig.~\ref{fig4}(a) surface plots of the correlation function
\begin{equation}
\gamma(j,t) = \langle\psi(t)\vert \sum_i c_i^\dagger c_i b_{i+j}^\dagger b_{i+j} \vert\psi(t)\rangle.
\label{gamma}
\end{equation}
$\gamma(j,t)$ enables us to follow the spread of bosonic degrees of freedom during the time evolution. At short times, $t\lesssim 5$ and small intensities,  $\gamma(j,t)\lesssim 0.1$, we observe a similar spread of the front of bosonic excitations away from the electron position at $j=0$. This spread is given by the Lieb-Robinson's  velocity (i.e., the maximal speed at which information propagates) that in both models equals  the maximal velocity of a free electron $v_\mathrm{LR}=2t_0$. Since bosons of both types are dispersionless, the observed velocity is due to the electron that moves away from the bosonic excitation.  At later times and larger intensities we observe a qualitative difference in $\gamma(j,t)$ between both models.    In the HM  there is an excess  of extra phonon excitations at and in the close proximity of the electron's position.  In contrast,  the HCM case bosonic degrees of freedom spread much more uniformly throughout the entire system. This represents the main mechanism that causes substantial slowing down of  relaxation in HCM at larger values of $\Delta$. As $\Delta$ increases far beyond the typical bosonic frequency, $\Delta\gtrsim \omega_0$, in a semiclassical  picture, the charge has to to travel ever  larger distance to dispose off the excess energy. 

\begin{figure} 
\includegraphics[width=0.41\textwidth]{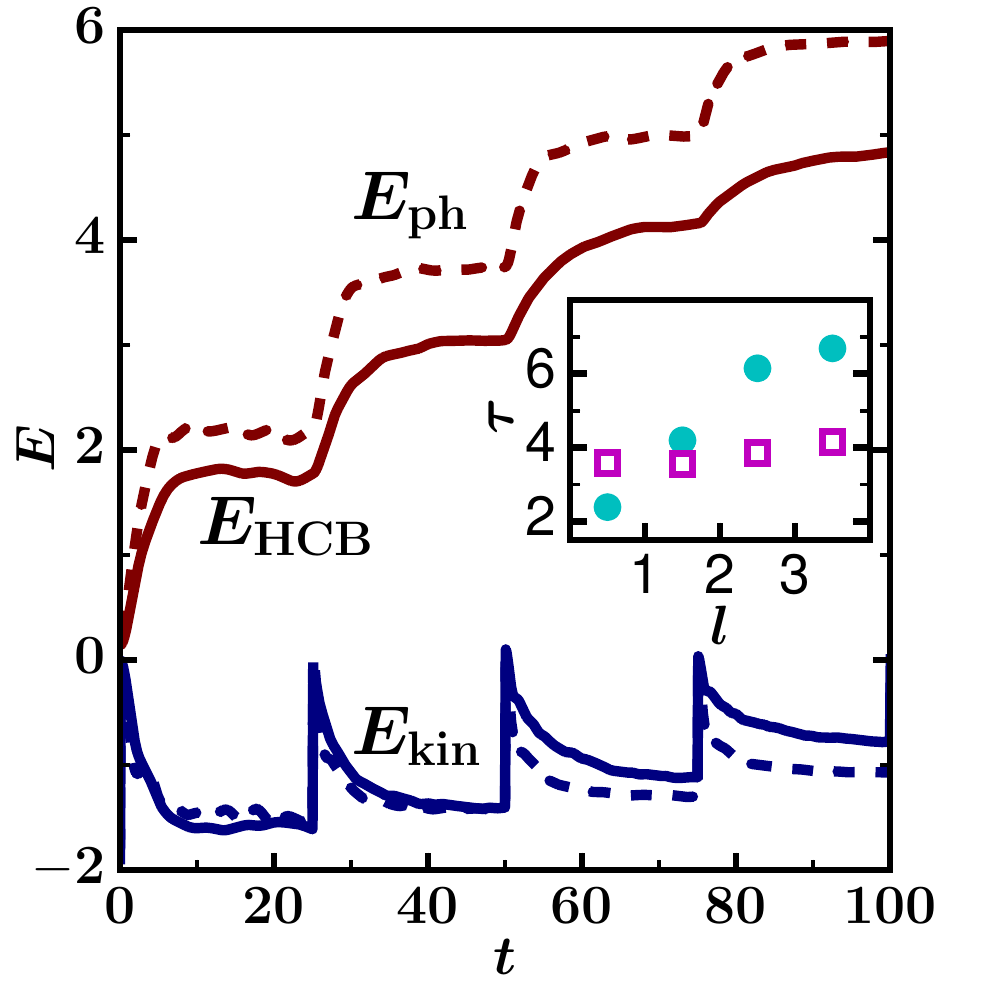}
\caption{  Kinetic energy $E_\mathrm{kin}$, phonon energy $E_\mathrm{ph}$ and hard core boson energy $E_\mathrm{HCB}$ vs time. Dashed (full) lines represent results for the HM (HBM). {$\omega=g=0.75$, and $L=16$  was used in this particular case. Note that the upper limit of   $E_\mathrm{HCB}$, reached at infinite temperature,  is $E_\mathrm{HCB}^\mathrm{max}=L\omega /2 = 6$.} Multiple spikes in $E_\mathrm{kin}$ are due to step-like jumps in the  phase $\phi(t) = \pi/2 \sum_{l=0}^3 \theta(t-l\Delta t)$ where $\Delta t = 25$ and $\theta(x)$ is the Heaviside step function.  In the insert we present relaxation time $\tau$ with squares (full circles) for the HM (HBM), extracted from the relaxation of $E_\mathrm{kin}(t)$ in distinct time intervals between successive step-like changes of $\phi(t)$.  
}
\label{fig5}
\end{figure}

We have demonstrated that the HC effects slow down the relaxation of highly exited charge carriers.  
Therefore, the future experiments showing the relaxation time vs. the excitation energy may shed light on the type of bosons, which are most strongly coupled to the carriers.
However,  in the majority of the experimental setups it is easy to tune the density of photoexcited carriers (e.g., by changing the fluence) but not necessarily  their energy. 
Direct numerical  simulations of the former problem cannot be carried out within the present model with a single charge carrier.  However, we expect that a qualitatively similar 
picture may  be  obtained from the studies of a slightly  different experimental setup when a  charge carrier  is driven by multiple pulses. The discussed scenario should hold independently of whether   the bosons are excited by a single or various charge carriers. At the end, what  matters for the hard--core effects is the density of the bosons and not their source. 

In Fig.~\ref{fig5} we show how the relaxation changes upon applying subsequent pulses to HM and HCM. 
{We have simulated this case by starting the  time evolution  from the polaron ground state wavefunction at $t=0^-$, followed by successive step-like jumps in the phase $\phi(t)$ as described in the caption of Fig.~\ref{fig5}. Each jump in $\phi(t)$ causes an abrupt jump of $E_\mathrm{kin}$  followed by a relaxation process in which  a decrease in $E_\mathrm{kin}$ is followed by the  increase of the corresponding  boson energy. 
In the HCM, the relaxation after succeeding pulses becomes visibly slower as   the density of  excited hard--core bosons becomes comparable with its value at infinite temperature, {\it i.e.}  $E_\mathrm{HCM}^\mathrm{max}/L= \omega/2$, see also the caption in Fig.~\ref{fig5} }
In contrast,  relaxation in the HM  does not show any substantial dependence on the number of  preceding  pulses. This is clearly seen in the   insert of   Fig.~\ref{fig5} where we present relaxation times for both models as extracted from the exponential fits to $E_\mathrm{kin}$ in distinct time intervals.

{\it In summary}, we propose a simple mechanism to distinguish between two different classes of   bosonic excitations that are responsible for the primary (fastest) relaxation mechanism of a photo excited charge carrier in time-resolved pump-probe experiments. The proposed mechanism is based on  the recognition that phonon degrees of freedom can  absorb essentially unlimited amount of energy while in contrast, the hard core effects, typical for {\it e.g.} magnon excitations, pose strict limits on the density of absorbed energy.  For this reason the relaxation dynamics of the charge carrier coupled to phonon degrees of freedom  very weakly depends on the excitation energy, while the opposite is true when the charge carrier is coupled to HC boson excitations.  In the latter case the relaxation becomes less effective when the absorbed energy approaches  the typical HC boson frequency, {\it i.e.} $\Delta\sim\omega$.  The density of excited bosons can be tuned either by changing the fluence of a single pump--pulse, or by driving the system by multiple pulses. In the latter case  the time span between the first and the last pulses should be significantly smaller than the secondary relaxation time when other, weakly coupled degrees of freedom
start to influence the relaxation process.

To gain clear distinction between the two different cases, we performed simulations on nearly identical models containing either Einstein phonons or dispersionless HC bosons. However, in more realistic  systems  dispersionless HC bosons should be replaced by, {\it e.g.},  dispersive magnetic excitations with a given magnon velocity $v_\mathrm{mag}$.  Still, even in this case the slowing down of relaxation due to HC effects is expected in two spacial dimensions when $\Delta\sim \omega (v_\mathrm{mag}\tau)^2$.
Up to our best knowledge,  the predictions of this work have not been yet tested experimentally. However, the very recent experiments on cuprates  show significant fluence--dependence of the relaxation times obtained for the electron--states which are far above the Fermi energy (see Fig. 5 in Ref. \cite{rameau15}). This result suggests that in cuprates the charges are most strongly-coupled to 
HC bosonic excitation, which are most probably of the magnetic origin.

\begin{acknowledgments}
We acknowledge stimulating discussions with U. Bowensiepen,  C. Giannetti. 
J.B. acknowledges discussions with A. Polkovnikov and M. Rigol as well as  the  support by the P1-0044 of ARRS, Slovenia.
M.M. acknowledges support from the DEC-2013/11/B/ST3/00824 project of the Polish National Science Center.
J.K. acknowledges financial assistance by the Alexander von Humboldt Foundation.
This work was performed, in part, at the Center for Integrated Nanotechnologies, a U.S. Department of Energy, Office of Basic Energy Sciences user facility. 
\end{acknowledgments}

\bibliographystyle{biblev1}
\bibliography{Polarons}

\end{document}